\documentclass[11pt]{article}
\input{psfig}
\input epsf
\epsfverbosetrue
\def\styleversion{1.002}\def\styledate{7 Jul 1995}
\def\titlepage#1#2{\clearpage%
\setcounter{footnote}{0}\pagestyle{empty}%
\mbox{\begin{tabular}{llr}
\epsfxsize=20mm
\epsfbox{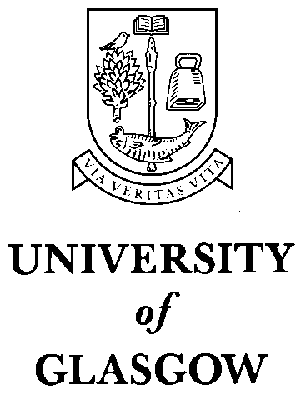}
& \begin{tabular}{c} 
\vspace*{-2.7cm}\\
{\Large Department of Physics \& Astronomy} \\
\multicolumn{1}{c}{\large Experimental Particle Physics Group} \\
{\small Kelvin Building, University of Glasgow,} \\
{\small Glasgow, G12 8QQ, Scotland} \\
{\small Telephone: +44 (0)141 339 8855 Fax: +44 (0)141 334 9029}\\
\end{tabular} 
& \begin{tabular}{r}
\vspace*{-5cm}\\
{\bf #1} \\
{#2}\\
\end{tabular}
\end{tabular}}

\vspace{2cm}

}

\def\collaboration#1{\vskip1em\begin{center}\it#1\end{center}}
\def\conference#1{\vskip1em\begin{center}\it#1\end{center}}

\def\title#1{\vskip1em\begin{center}\large\bf#1\end{center}\vskip1.5em}
\def\abstract{\begin{center}{\bf Abstract}\\[\baselineskip]%
\end{center}\quotation\small}

\def\endtitlepage{
\setcounter{footnote}{0}\let\titlepage\relax\vfill
\newpage\setcounter{page}{1}\pagestyle{plain}\pagenumbering{arabic}%
\gdef\@thanks{}\gdef\@author{}\gdef\@title{}\let\thanks\relax}
\def\Journal#1#2#3#4{{#1} {\bf #2}, #3 (#4)}

\def\NPB{{\em Nucl. Phys.} B}
\def\PLB{{\em Phys. Lett.}  B}
\def\PRL{\em Phys. Rev. Lett.}

\def\ZPC{{\em Z. Phys.} C}

\def\be{\begin{equation}}
\def\ee{\end{equation}}
\def\bea{\begin{eqnarray}}
\def\eea{\end{eqnarray}}
\begin{document}
\bibliographystyle{unsrt}    

\begin{titlepage}{GLAS--PPE/97--01}{\today}

\title{Recent Results from the study of QCD and Jets at HERA}
\centerline{L.E. Sinclair\footnote{e-mail: sinclair@desy.de}}
\collaboration{for the ZEUS and H1 Collaborations}
\conference{Talk presented at the Lake Louise Winter Institute,
Particles and the Universe,\\ Lake Louise, Alberta, Canada, 
February 1997.}

\begin{abstract}
The HERA experiments H1 and ZEUS have recently presented a number of
interesting results from studies of the hadronic final state 
produced in $ep$ collisions.  These studies have focussed on deep 
inelastic scattering events and on the photoproduction of jets, and 
also on the comparison of the final states produced in these
two regimes.  Some of these results are reviewed here.
\end{abstract}
\end{titlepage}

\section{Introduction}

Analyses of the distribution of final state 
hadrons\footnote{In the $ep$ world it is common to refer to all of the
final state particles other than the scattered lepton as the 
``hadrons''.}
in $ep$ scattering
events can provide tests of the theory of strong interactions,
quantum chromo-dymamics (QCD).
These studies have yielded information on the value and running of 
$\alpha_s$, 
on the physics of parton shower evolution, on 
the  structure of the photon and proton and on the
underlying parton scattering dynamics.
There are too many recent results from H1 and ZEUS 
to attempt a complete review here.  Instead I will concentrate
on the measurement of fragmentation functions
in deep inelastic scattering (DIS)~\cite{fragf94_h1}, some
results from the measurement of jet cross sections in 
photoproduction~\cite{jets_h1,dij_zeus}, a study of prompt photon
production~\cite{prompt_zeus} and some results from the
comparison of the hadronic final state produced in deep inelastic 
scattering with that produced in 
photoproduction~\cite{etflow_h1,shapes_zeus}.  These results pertain
to data collected in the 1994 and 1995 periods of data taking,
during which HERA provided collisions of 27.5~GeV positrons against 
820.0~GeV protons, delivering about 3~nb$^{-1}$ and 6~nb$^{-1}$
of luminosity respectively.

\section{Hadronic Final State in Deep Inelastic Scattering}

In the leading order quark parton model of DIS,
the incident positron scatters from a quark in the proton via the
exchange of a boson whose negative invariant mass squared is
denoted $Q^2$.  The quark carries a fraction $x_{Bj}$ of the 
proton's longitudinal momentum and energy into the hard subprocess.
This simple picture is sufficient to allow the appreciation of the 
results presented here.

\subsection{Fragmentation Functions}

To characterize the parton showering and subsequent hadronization
which determine the distribution of hadrons in the final state,
the distribution of the scaled momentum $x_p$ is measured.  The
scaled momentum is defined as the magnitude of a hadron's momentum 
relative to the maximum momentum available to it,
$x_p = p_{\mbox{\scriptsize HADRON}} / p_{\mbox{\scriptsize MAX}}$.
  The measurement of scaled momentum
distributions is carried out in the ``Breit'' or ``brickwall'' frame
where $p_{\mbox{\scriptsize MAX}} = - Q / 2$.  In this frame the 
current hemisphere in
DIS is directly comparable to one hemisphere in
an $e^+e^-$ annihilation event where the centre-of-mass energy
$E^*$ would represent the energy scale, $Q$, and 
$p_{\mbox{\scriptsize MAX}} = E^* / 2$.

\begin{figure}[h]
\begin{center}
\leavevmode
\psfig{figure=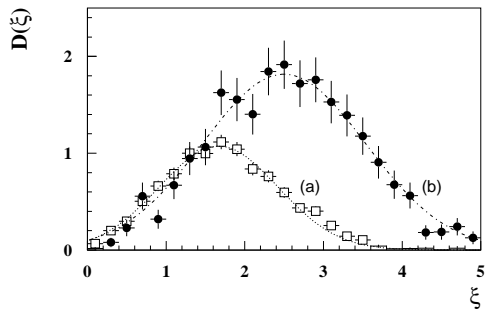,height=1.7in}
\end{center}
\caption{Fragmentation functions for 
(a) $12 < Q^2 < 80$~GeV$^2$ and 
(b) $Q^2 > 100$~GeV$^2$.  Gaussian fits to the distributions are 
superimposed.
\label{fig:fragf0_h1}}
\end{figure}

In fact a variable transformation is commonly made to
$\xi = \ln (1 / x_p)$ which has an approximately Gaussian distribution.
The distribution of $\xi$ is called a fragmentation function.
As an example, fragmentation functions in two bins of $Q^2$ are shown
in Figure~\ref{fig:fragf0_h1}
for data taken by H1 during
the 1993 running period~\cite{fragf93_h1}.
The data support the intuitive expectation that as $Q^2$ increases,
the average charged multiplicity should increase, $\langle x_p \rangle$
 should  decrease, and $\langle \xi \rangle$ should increase.

\begin{figure}[h]
\begin{center}
\leavevmode
\psfig{figure=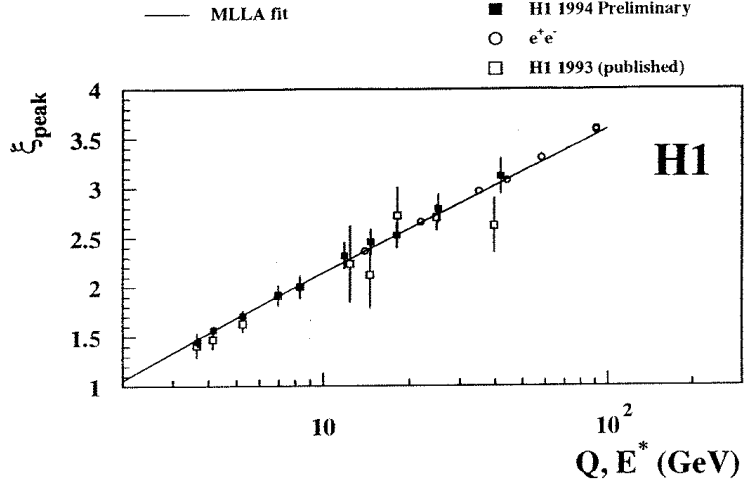,height=2.5in}
\end{center}
\caption{Peak values of the Gaussian fit to the $\xi = \ln (1 / x_p)$
distribution versus $Q$ for H1 1993 (open squares) and H1 1994 
(filled squares) data.  The line shows a linear fit to the data.
Also shown is the corresponding $\xi$
distribution parameter versus $E^*$ for $e^+e^-$ data (circles).
\label{fig:fragf3_h1}}
\end{figure}

To quantify the $Q^2$ evolution of $\ln (1 / x_p)$, a Gaussian fit of
the $\xi$ distribution has been made for several different $Q$ bins.
The resulting fit parameters for the peaks of the Gaussian
distributions, $\xi_{\mbox{\scriptsize peak}}$, are shown in 
Figure~\ref{fig:fragf3_h1}~\cite{fragf94_h1}.
The data are consistent with a QCD inspired picture of fragmentation 
in which $\xi_{\mbox{\scriptsize peak}}$ is expected to scale 
approximately linearly with $\ln Q$.  
A pure phase space model would entail a slope of unity in the 
dependence
of $\xi_{\mbox{\scriptsize peak}}$ on $\ln Q$ while a fit to the data 
yield $0.75 \pm 0.05$.
Therefore the data prefer coherent fragmentation as predicted by 
QCD.  (See also reference~\cite{fragf93_h1} and references therein.)

Results for the evolution of the $\xi_{\mbox{\scriptsize peak}}$ 
fit parameter with $E^*$ for
$e^+e^-$ data are also shown in Figure~\ref{fig:fragf3_h1}.  The
consistency of the $e^+e^-$ data with the $ep$ data supports the 
application of a single universal treatment of parton showering and 
hadronization.

\section{Photoproduction of Jets}

\begin{figure}[h]
\begin{center}
\leavevmode
\psfig{figure=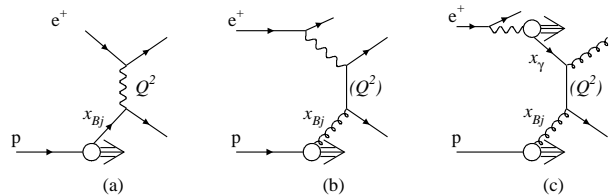,height=1.in}
\end{center}
\caption{Diagrams for DIS (a) and direct (b) and 
resolved (c) photoproduction.
\label{fig:HPP}}
\end{figure}

Deep inelastic scattering processes (Figure~\ref{fig:HPP}(a)) are
characterized by rather high values of the negative squared four 
momentum of the boson (usually a photon), 
$Q^2 > $ $\mathcal{O}$$(1)$~GeV$^2$.
In photoproduction on the other hand the photon is nearly real.
When jets are produced in photoproduction, their transverse energy
sets the energy scale of the process, also often represented by $Q^2$.

At leading order there are two photoproduction processes which lead to
the production of high transverse energy jets.  In ``direct'' 
photoproduction the photon participates directly in a hard scattering 
with a parton from the proton as illustrated in 
Figure~\ref{fig:HPP}(b).  In ``resolved'' photoproduction 
(Figure~\ref{fig:HPP}(c)) the photon
first fluctuates into a hadronic state a parton from which enters the
hard subprocess.  It is therefore necessary to introduce another
Bjorken scaling variable, denoted $x_{\gamma}$, which represents
the fraction of the photon's momentum which is carried by the 
interacting parton.

\subsection{Jet Cross sections}

\begin{figure}[h]
\begin{center}
\leavevmode
\psfig{figure=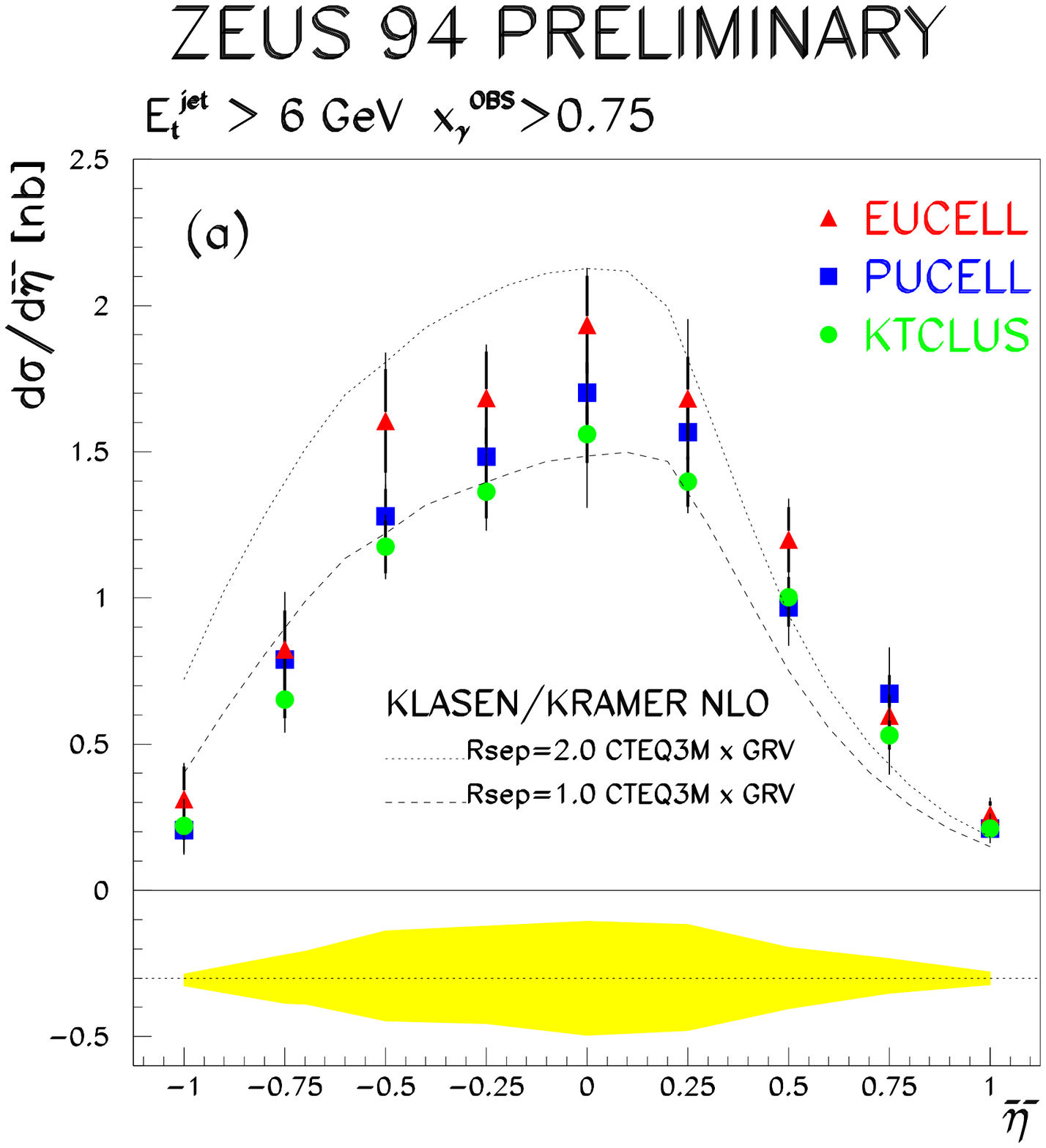,height=2.7in}\psfig{figure=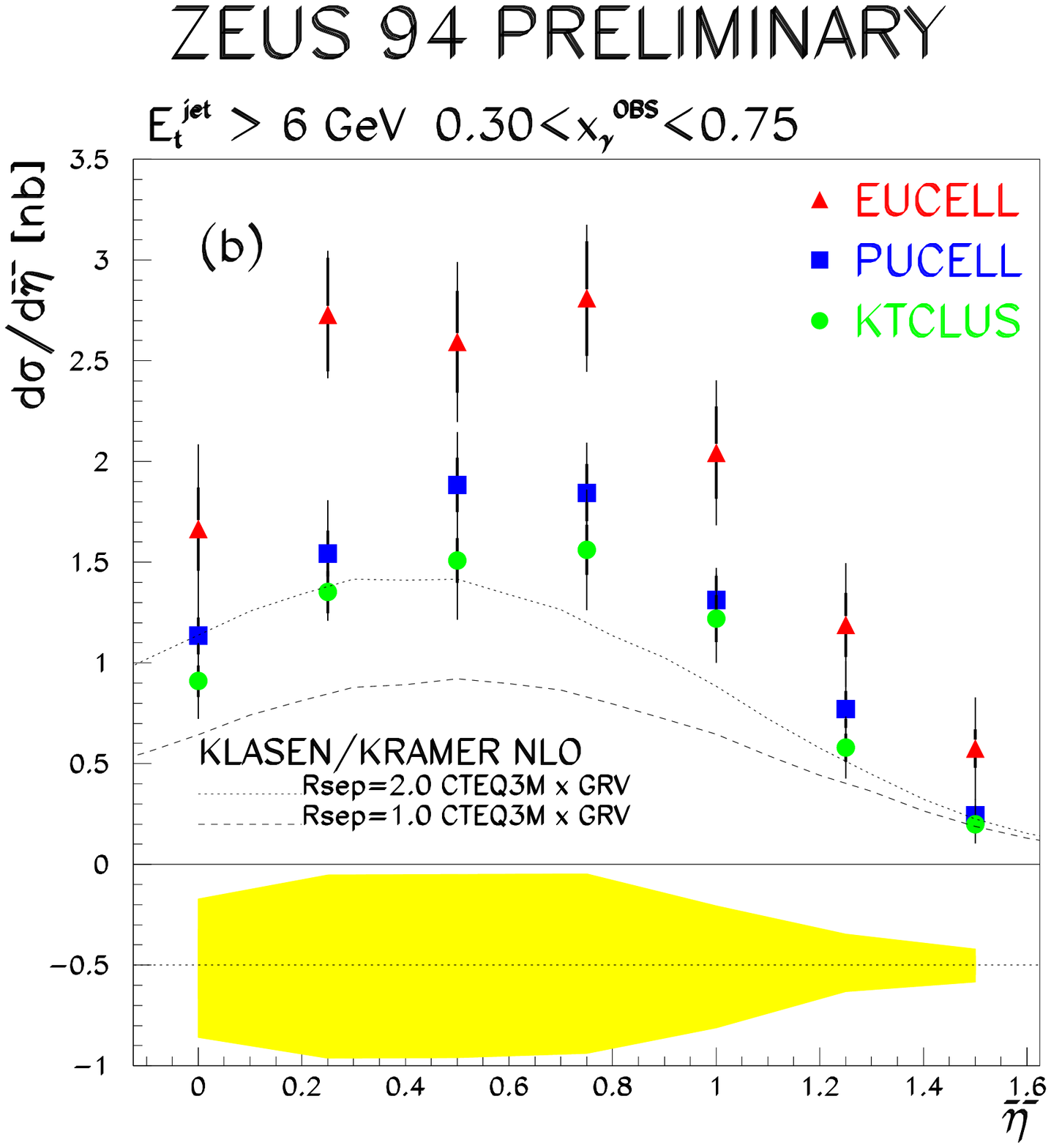,height=2.7in}
\end{center}
\caption{Dijet photoproduction cross section, $d\sigma / d\bar{\eta}$,
for (a) $x_{\gamma} > 0.75$ 
and (b) $0.30 < x_{\gamma} < 0.75$.  
The different symbols show the measured cross section using
three different jet finding algorithms.
The curves show NLO 
calculations of the cross section using the CTEQ3M proton parton 
distributions and the
GRV photon parton distributions for two different values
of the R$_{\mbox{\scriptsize sep}}$ parameter (see text).  
(The shaded band shows
the correlated uncertainty on the absolute energy scale of the 
detector.)
\label{fig:jets_hixg_zeus}}
\end{figure}

ZEUS has measured the dijet photoproduction cross section differential
in the average jet pseudorapidity, $d \sigma / d \bar{\eta}$, for 
events with at least two 
$E_T^{\mbox{\scriptsize JET}} > 6$~GeV jets~\cite{dij_zeus}.
By making the requirement that $x_{\gamma} > 0.75$,
we effectively examine the direct photoproduction cross 
section~\cite{dij93_zeus}.
The results are shown in Figure~\ref{fig:jets_hixg_zeus}(a)
where three different jet finders have been used to define the
cross section.
EUCELL and PUCELL are cone algorithms which 
satisfy the Snowmass convention~\cite{snow}.
The difference between them
illustrates the sensitivity of the measurements to the choice of
seed particle and the treatment of overlapping cones.
KTCLUS is a cluster jet finding algorithm~\cite{ktclus}.

The results are compared to next to leading order (NLO)
predictions for the cross sections using the
KTCLUS algorithm and two different values of the 
R$_{\mbox{\scriptsize sep}}$ 
parameter~\cite{klasen}.
(R$_{\mbox{\scriptsize sep}}$ defines the farthest distance in
pseudorapidity and azimuth that two partons can have from one another 
and still be combined into one jet~\cite{rsep}.)
The prediction shown by the dashed line may be directly compared with
either the PUCELL or KTCLUS cross sections which are equivalent at
NLO and have an effective
R$_{\mbox{\scriptsize sep}}$ value of 1.  
The theory provides a reasonably good 
description of both the normalization and shape of the measured
distribution.
In EUCELL the effective value of R$_{\mbox{\scriptsize sep}}$ is not
precisely defined but is in the range 1.5$\rightarrow$2.
The theory is
therefore able to describe the spread of the cross sections due
to the different jet finding algorithms.

The situation is considerably worse for the resolved photon processes.
The cross section $d\sigma / d\bar{\eta}$ for 
$0.30 < x_{\gamma} < 0.75$ is shown in 
Figure~\ref{fig:jets_hixg_zeus}(b).
Here the NLO calculation for the KTCLUS algorithm (dashed line) does 
not describe the measured KTCLUS cross section.
Also the cross section suffers more variation under change of jet
algorithm and this variation is not describable by the NLO calculation.
However it must be noted that much of the uncertainty due to the 
photon structure is contained in the resolved sample and 
that also the higher order processes which are not included in the
fixed order calculation would show up at low $x_{\gamma}$.

\begin{figure}[h]
\begin{center}
\leavevmode
\psfig{figure=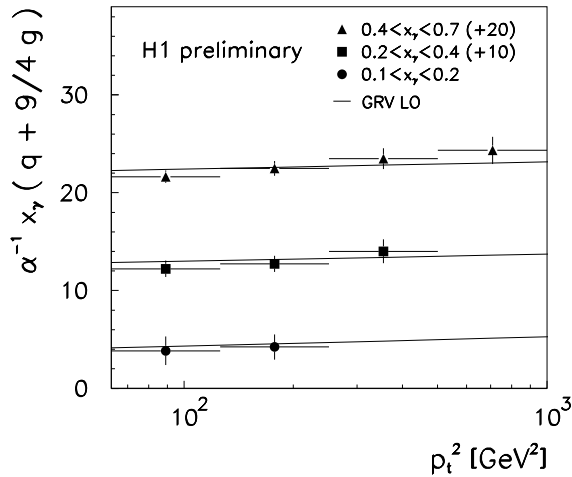,height=2.5in}
\end{center}
\caption{The symbols show the photon parton density versus $p_t^2$ in 
three bins of $x_{\gamma}$.  The curves represent the photon parton 
density calculated using the GRV-LO parton distribution functions.
\label{fig:jets_pdf_h1}}
\end{figure}

It is possible to define an effective parton distribution of the
photon, $f_{i/{\gamma}} = \sum (q + \bar{q}) + 9/4 g$, which can
be measured in photoproduction~\cite{gpdf}.  
Note that in contrast to deep
inelastic $e\gamma$ experiments which only determine the quark
content of the photon, the HERA $\gamma p$ measurements
are sensitive also to the gluonic term of $f_{i/{\gamma}}$.  By
comparison of measured parton level cross sections with the
PYTHIA~\cite{PYTHIA} Monte Carlo leading order prediction H1 have 
extracted the effective parton distribution of the 
photon in three bins of $x_{\gamma}$ as a function of
the hard scattering scale, $p_t^2$ (the square of the partonic 
transverse momentum).~\cite{jets_h1}  This is shown in 
Figure~\ref{fig:jets_pdf_h1}.

The data are consistent with a logarithmic rise of $f_{i/{\gamma}}$
with $p_t^2$.  We recall that a calculation of the quark content
of the photon from the box diagram 
$\gamma \gamma^* \rightarrow q \bar{q}$ yields the form
$F_2^{\gamma \mbox{\small{~box}}} = a(x_{\gamma}) + b(x_{\gamma}) \log p_t^2$~\cite{witten}.
It is therefore interesting to speculate whether the logarithmic
rise of $f_{i/\gamma}$ with $p_T^2$ reveals a component of the
resolved photon which results directly from a splitting of the photon
into a $q\bar{q}$ pair.  This has been called the ``anomalous'' 
component in contrast to that part of the photon's structure which
behaves as though the photon has fluctuated into a vector meson.

\subsection{Prompt Photon Production}

A particularly interesting photoproduction event topology occurs
when a high transverse energy photon takes the place of one of the
jets.  Such processes are called prompt photon processes.  These
events provide a clean QCD laboratory for the study of quark jet
showering and fragmentation and for the measurement of the quark
content of the resolved photon.  This is also the best process for 
the study of initial-final state QCD coherence.  In direct 
photoproduction there is a single flow of colour from the incoming to 
the outgoing quark, in contrast to the situation in generic dijet 
production where many different possible colour flows contribute to 
each event topology.

\begin{figure}[h]
\begin{center}
\leavevmode
\psfig{figure=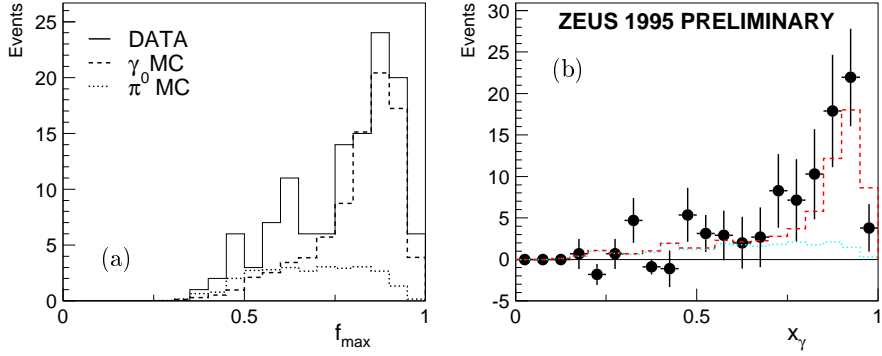,height=1.8in}
\end{center}
\caption{(a) The distribution of $f_{\mbox{\scriptsize max}}$ for the data (solid 
histogram), the simulated prompt photon events (dashed histogram) and 
the simulated $\pi^0$ background (dotted histogram).
In (b) the black dots show the $x_{\gamma}$ distribution for prompt 
photon events after background subtraction.  The dashed histogram 
shows the expectation from the PYTHIA Monte Carlo simulation.  The 
dotted 
histogram shows the simulated $x_{\gamma}$ distribution separately 
for the (leading order) resolved component.
\label{fig:prompt_zeus}}
\end{figure}

To select a sample of prompt photon events a jet and an isolated 
electromagnetic energy deposit in the ZEUS barrel calorimeter with 
no associated 
charged particle found in the tracking chamber, are required.  However
some dijet events, where one jet consists largely of a single 
$\pi^0$ or $\eta$, can pass this selection.  This contribution can be 
understood
using a Monte Carlo simulation as shown in
Figure~\ref{fig:prompt_zeus}(a)~\cite{prompt_zeus}.
The fraction of the electromagnetic shower energy which
is contained in the highest energy calorimeter cell is denoted
$f_{\mbox{\scriptsize max}}$.  Photons give signals with
higher values of $f_{\mbox{\scriptsize max}}$ than the $\pi^0$ 
background
and it is possible to describe the measured distribution of 
$f_{\mbox{\scriptsize max}}$
by allowing for a small $\pi^0$ component.  This also allows for
an admixture of $\eta$'s.

The $x_{\gamma}$ distribution for prompt photon events after 
subtraction of the $\pi^0$ background is shown in 
Figure~\ref{fig:prompt_zeus}(b).  The distribution is roughly
consistent in shape and normalization with a leading order QCD 
prediction.  Most of the events have high values of $x_{\gamma}$ and
are therefore associated with direct prompt photon production.  We
look forward
to a larger sample of prompt photon events with the 1996 and 1997
delivered luminosities.

\section{Comparison of Deep Inelastic Scattering and Photoproduction}

Two interesting recent results 
compare the hadronic final states in deep inelastic scattering
and photoproduction.  These studies are particularly appealing to the
experimentalist since it is possible to make some interpretation of 
the 
data which does not rely on a comparison of the data to theory.

\subsection{Transverse Energy Flow}

\begin{figure}[h]
\begin{center}
\leavevmode
\psfig{figure=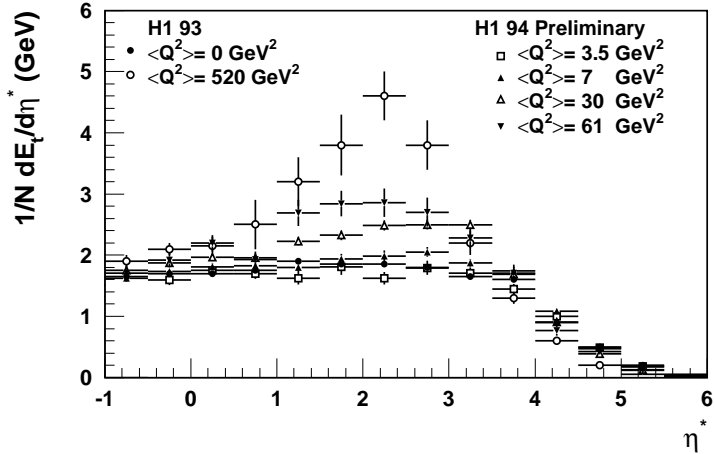,height=2.5in}
\end{center}
\caption{Average transverse energy, $E_T$, in bins of the 
hadronic centre of mass frame pseudorapidity, $\eta^*$, and 
$\langle Q^2 \rangle$.
\label{fig:etflow1_h1}}
\end{figure}

H1 has measured the flow of transverse energy in the hadronic centre
of mass frame across a large range of $Q^2$, as shown in 
Figure~\ref{fig:etflow1_h1}~\cite{etflow_h1}.
The hadronic centre of mass frame pseudorapidity, $\eta^*$, is defined
such that the proton direction is negative so the current region is
at positive $\eta^*$.

The $E_T$ in the current region
of course grows with $\langle Q^2 \rangle$.  However the remarkable 
feature of this distribution is that the mean transverse energy in 
the central region, $\eta^* \sim 0$, is practically independent of 
$Q^2$ from the soft photoproduction region, $\langle Q^2 \rangle = 0$,
up to the very deep inelastic region,
$\langle Q^2 \rangle = 520$~GeV$^2$.
H1 therefore find that there is a smooth transition in the
development of the global flow of transverse energy between the
photoproduction and DIS regimes.

\subsection{Jet Shapes}

A complementary result from ZEUS concerns the internal structure
of jets~\cite{shapes_zeus}.
The jets are found using a cone algorithm with jet radius $R = 1$.
The jet shape parameter, $\psi(r)$, has been defined as the fraction of
the jets transverse energy which is contained within a smaller 
concentric cone of radius $r$.
\begin{figure}[h]
\begin{center}
\leavevmode
\psfig{figure=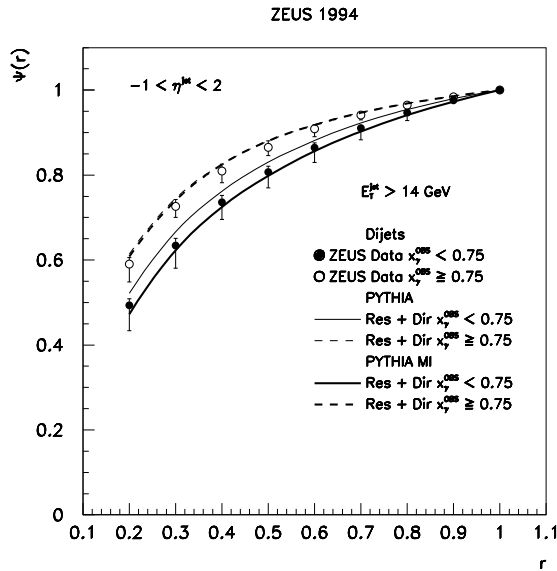,height=3.in}
\end{center}
\caption{Jet shapes in dijet photoproduction for high (open circles) 
and low (black dots) $x_{\gamma}$ samples.  The PYTHIA predictions for
the low and high $x_{\gamma}$ regions are shown by the solid and dashed
lines respectively.  Thicker lines show the PYTHIA predictions where 
multiple interactions between the constituents of the photon and 
proton have been allowed.
\label{fig:shapesgp9bw_zeus}}
\end{figure}
  This is shown in 
Figure~\ref{fig:shapesgp9bw_zeus} for a sample of
photoproduction events which have two jets of 
$E_T^{\mbox{\scriptsize JET}} > 14$~GeV and $-1 < \eta^{\mbox{\scriptsize JET}} < 2$.
The jet shape is shown separately for direct and resolved photon 
events.  The jet shape for the $x_{\gamma} \geq 0.75$
sample is narrower than the shape of the jets for the resolved photon
sample.  This effect is reproduced by the parton shower Monte Carlo
model PYTHIA.  Allowing for secondary interactions between the 
photon 
and proton remnant particles in the simulation improves
the description of the $x_{\gamma} < 0.75$ data.

\begin{figure}[h!]
\begin{center}
\leavevmode
\psfig{figure=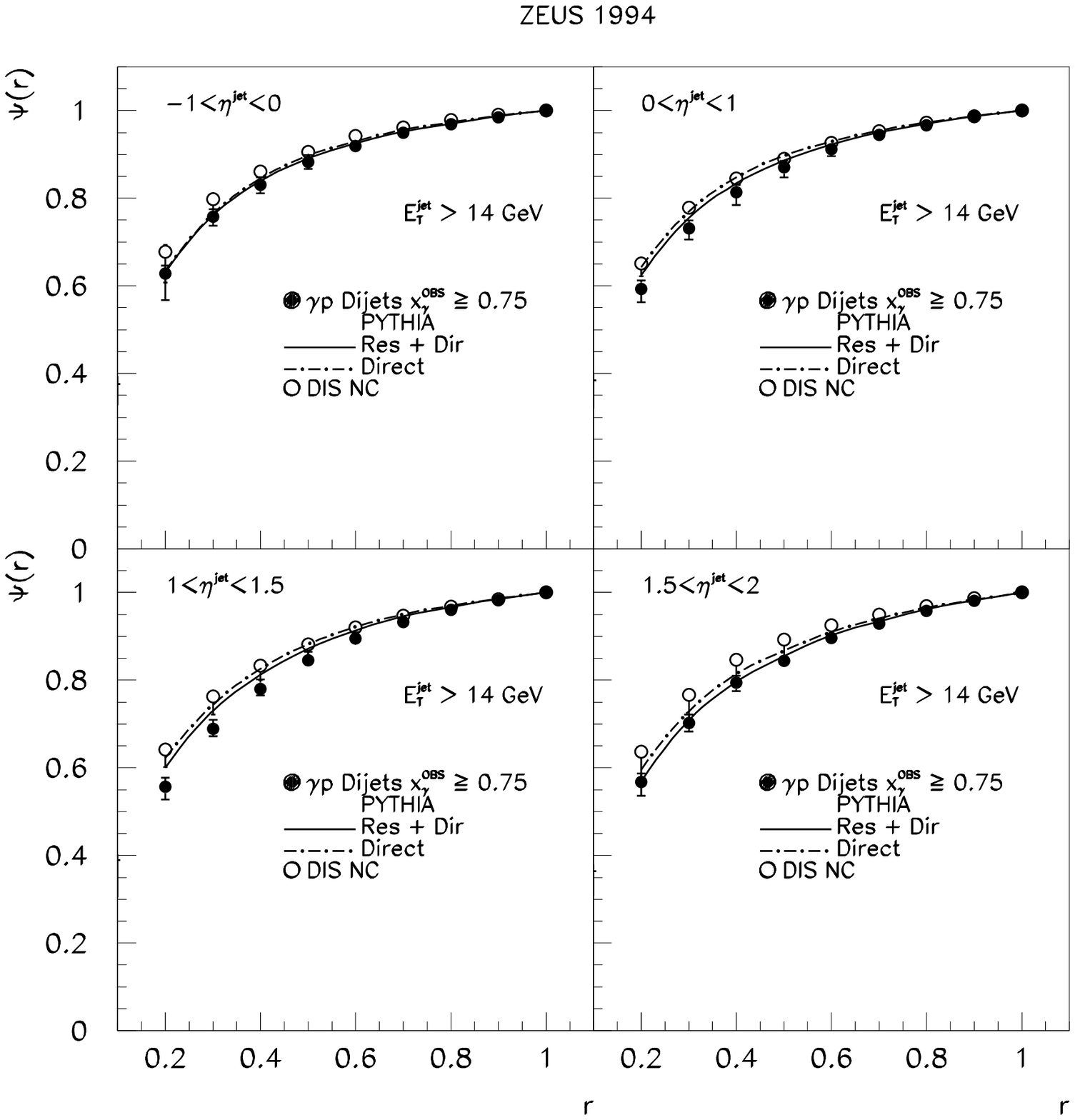,height=4.in}
\end{center}
\caption{Jet shape in direct photoproduction (dark dots) compared
with jet shape in DIS (light dots) in four 
bins of $\eta^{\mbox{\scriptsize JET}}$.  The data are not corrected for detector effects
and are compared with PYTHIA Monte Carlo predictions to which a full
simulation of the ZEUS detector has been applied.
The prediction for the the high $x_{\gamma}$ jet shape is shown by
the dark line and the prediction for the pure leading order direct 
processes is shown by the light line.
\label{fig:shapes61_zeus}}
\end{figure}

One can speculate that the observed difference in the high and
low $x_{\gamma}$ jet shapes is due primarily to
the colour charge of the parton which initiates the jet.  In direct
photoproduction this parton is predominantly a quark whilst in resolved
photoproduction the jets are much more often initiated by gluons.
If the direct photoproduction jets are narrower than the resolved
photoproduction jets because of the predominance of quark jets, then
one would expect the jet shapes of the high $x_{\gamma}$ events to be 
comparable to jet shapes in DIS.  This is
indeed the case as shown in 
Figure~\ref{fig:shapes61_zeus}~\cite{shapes_zeus}.
The high $x_{\gamma}$ jet shape
is shown in four bins of $\eta^{\mbox{\scriptsize JET}}$ and is found to be comparable
with the jet 
shape for DIS events ($Q^2 > 4$~GeV$^2$) having
at least one $E_T > 14$~GeV jet~\cite{shapes_zeus}.

\section{Conclusions}

Quantum chromodynamics is a successful theory.  For a number of years
it has been clearly
established that in the perturbative regime QCD matrix elements
describe the dynamics of parton scattering.  
Here it is shown that $ep$ scattering provides the means
to perform detailed tests of the QCD inspired models for parton 
showering and hadronization.  These models, as implemented in Monte 
Carlo event generators, are able to describe the measured observables.
There is also strong evidence
that the fragmentation process has a universal 
nature given the similarity of the hadronic final states in
$e^+e^-$ and in DIS events as well as in DIS and photoproduction 
events.

\section*{Acknowledgments}
My thanks to the members of the H1 and ZEUS collaborations who
have performed the analyses discussed here.
The organizers of the Lake Louise Winter Institute are to be
congratulated for having succeeded to provide a most enjoyable
and stimulating atmosphere for the exchange of ideas.


\end{document}